\documentclass[a4paper,12pt]{article}
\usepackage{latexsym,amssymb}
\usepackage{amsmath,amssymb,amsxtra,amscd,amsthm}
\usepackage[mathscr]{eucal}

\DeclareFontFamily{U}{rsfs}{\skewchar\font127 }
\DeclareFontShape{U}{rsfs}{m}{n}{
   <5> rsfs5
   <6> rsfs6
   <7> rsfs7
   <8> rsfs8
   <9> rsfs9
   <10> rsfs10
   <10.95> rsfs11
   <12> rsfs12
   <14.4> rsfs14
   <17.28> rsfs17
   <20.74> rsfs20
   <24.88> rsfs25
   <29.86-> rsfs30}{}
\DeclareMathAlphabet\scr{U}{rsfs}{m}{n}

\def\mZ{\mathbb{Z}}

\def\tL{\mathrm{L}}

\def\tS{\mathrm{S}}

\newfont{\HUGE}{cmssbx10 scaled 4000}

\def\bb1{\textup{\small{1}} \kern-3.8pt \textup{1}}

\def\SL2Z{\tS\tL(2,\mZ)}

\numberwithin{equation}{section}

\providecommand{\href}[2]{#2}

\newtheoremstyle{plain} 
  {5pt}
  {10pt}
  {\rmfamily} 
  {}
  {\scshape} 
  {}
  {\newline} 
  {}
\swapnumbers
\theoremstyle{break}

\def\a{\alpha}

\def\l{\lambda}

\newcommand{\wip}{{\sf w}^+}
\newcommand{\wim}{{\sf w}^-}
\newcommand{\wipm}{{\sf w}^\pm}
\newcommand{\wimp}{{\sf w}^\mp}

\begin{document}
\begin{titlepage}
\begin{flushright}
DISTA-2008\\
hep-th/yymmnnn
\end{flushright}
\vskip 1.5cm
\begin{center}
{\LARGE \bf
Pure Spinors for General Backgrounds$^\dagger$
}
\vfill
{\large Pietro Fr\'e$^{~a}$ and Pietro Antonio Grassi$^{~b}$} \\
\vfill {
$^{a}$ DISTA, University of Eastern Piedmont, v. Bellini 25/g, \\
 15100 Alessandria, Italy, and INFN - Torino, Italy\\
\vskip .2cm
$^{b}$
Department of Theoretical Physics, University of Torino, v. Giura, 1, Torino, Italy
}
\end{center}
\vfill
\begin{abstract}
We show the equivalence of the different types of pure spinor constraints geometrically derived
from the Free Differential Algebras of $\mathcal{N}=2$ d=10 supergravities. Firstly, we compute the general solutions
of these constraints, using both a $\mathrm{G}_2$ and an $\mathrm{SO(8)}$ covariant decomposition
of the 10d chiral spinors. Secondly, we verify that the number
of independent degrees of freedom is equal to that implied by the Poincar\'e pure spinor constraints so-far used for
superstrings, namely twenty two.
Thirdly, we show the equivalence between the FDA type IIA/B constraints among each other and with the Poincar\'e ones.

\end{abstract}
\vfill
\vfill
\vspace{1.5cm}
\vspace{2mm} \vfill \hrule width 3.cm {\footnotesize $^ \dagger $
This work is supported in part by the European Union RTN contract
MRTN-CT-2004-005104 and by the Italian Ministry of University (MIUR) under
contracts PRIN 2005-024045 and PRIN 2005-023102}

\date{February 2008}
\end{titlepage}

\tableofcontents

\section{Introduction}

In order to have a constructive derivation of the pure spinor
sigma model on general supergravity backgrounds, we decided in \cite{marp} to derive it from the rheonomic approach
to supergravity. The latter is based on a Free Differential Algebra (FDA)
and the pure spinor formulation is based on the BRST extension of that FDA.
The closure of the BRST algebra leads to pure spinor constraints which look
different from those used on  \cite{Howe:1991bx,Howe:1991mf,Berkovits:2000fe}.
Therefore, it is crucial to prove the  equivalence of the new constraints with the old ones. A similar analysis was performed in a
previous paper \cite{Fre':2006es} and recently discussed also in a conference proceedings \cite{Fre:2008qw}.
\par
We discuss the pure spinor constraints as derived from FDA in the type IIA and
type IIB backgrounds. Let us name the solutions of such equations  the \textit{FDA pure spinors}. On the other hand
let us denote \textit{Poincar\'e pure spinors} those which solve the  constraints
so-far used for type II superstrings
\cite{Howe:1991bx,Howe:1991mf,Berkovits:2000fe} which read as
follows\footnote{The adopted name Poincar\'e refers to the fact that these constraints can be deduced by exploiting two
copies of the Poincar\'e Lie superalgebra in d=10, a left-handed and a right-handed one in the type IIA case and two independent left-handed ones
in the type IIB case. It is important to stress the word Lie superalgebra as opposed to the word FDA. Indeed the different
constraints which are naturally adapted to a generic supergravity background follow from the complete algebraic structure underlying
supergravity which is the FDA extension of the superPoincar\'e Lie algebra and which mixes the two spinor chiralities because of the Ramond-Ramond
$p$-forms }
\begin{equation}\label{pureflat}
\bar \l_1 \Gamma^{\underline a} \l_1=0\,, \hspace{2cm}
\bar\l_2  \Gamma^{\underline a} \l_2=0\,.
\end{equation}
We use the notation $\l_A$ (with $A=1,2$) for the pure spinors and we
distinguish between type IIA and IIB by choosing the chirality of $\l_2$.
\par
 Although the choice of  the Poincar\'e
constraints (\ref{pureflat}) is feasible  they imply  unconventional  superspace
constraints for supergravity.  Therefore  it becomes quite difficult
to construct explicitly the corresponding pure spinor sigma
 given a solution of the supergravity field equations. On the other hand,
the new pure spinor constraints  derived in \cite{marp}
from the FDA structure
are those naturally adapted to a generic  background and allow the immediate writing of the corresponding pure spinor string action on any
supergravity  on-shell configuration.
\par
Obviously, we need to show that these new constraints
lead to the correct amount of independent degrees of freedom
 to cancel the conformal central charge.
\par
The three situations,Poincar\'e, type IIA and IIB are summarized in the table 1.
{
\begin{table}
  \centering
 {\small
$$
  \vbox{
  \offinterlineskip
  \halign
    {&\vrule# &\strut\hfil #\hfil &\vrule# &\hfil #\hfil &\vrule#
     &\hfil #\hfil &\vrule# &\hfil #\hfil  &\vrule# \cr
   \noalign{\hrule}
      height5pt
      &\omit& &\omit& &\omit&
      \cr
      &\hbox to4 true cm{\hfill Poincar\'e \hfill}&
      &\hbox to4 true cm{\hfill type IIA  \hfill}&
      &\hbox to4 true cm{\hfill type IIB  \hfill}&
      \cr
      height5pt
      &\omit& &\omit& &\omit&
      \cr
    \noalign{\hrule}
      height5pt
      &\omit& &\omit& &\omit&
      \cr
      &T: ~~$\sum_A \overline\lambda_A \Gamma_{\underline{a}} \lambda_A =0$&
      &T: ~~$\sum_A \overline\lambda_A \Gamma_{\underline{a}} \lambda_A =0$&
      &T: ~~$\sum_A \overline\lambda_A \Gamma_{\underline{a}} \lambda_A =0$&  \cr
      height10pt
      &\omit& &\omit& &\omit&
      \cr
      &$\sum_A (-)^A \overline\lambda_A \Gamma_{\underline{a}} \lambda_A  =0 $&
       &$~\mathbf{B}_{[2]}: ~\sum_A (-)^A \overline\lambda_A \Gamma_{\underline{a}} \lambda_A  V^{\underline{a}}=0~~$&
        &$~\mathbf{B}_{[2]}: ~\sum_A (-)^A \overline\lambda_A \Gamma_{\underline{a}} \lambda_A  V^{\underline{a}}=0~~$&
       \cr
      height10pt
      &\omit& &\omit& &\omit&
      \cr
       &\omit&
       &$~\mathbf{C}_{[3]}: ~\overline\lambda_1 \Gamma_{[{\underline{a}} {\underline{b}}]} \lambda_2
       V^{\underline{a}} V^{\underline{b}}=0~~$&
       &$~\mathbf{C}_{[2]}: ~\overline\lambda_1 \Gamma_{\underline{a}} \lambda_2  V^a =0~~$&
      \cr
      height10pt
      &\omit& &\omit& &\omit&
      \cr
      &\omit&
 &$~C_{[1]}: ~\overline\lambda_1 \lambda_2   =0~~$& &\omit& \cr
      height10pt
      &\omit& &\omit& &\omit&
      \cr
      height5pt
      &\omit& &\omit& &\omit&
      \cr
    \noalign{\hrule} }
  }
$$
}
 \label{fieldcont2}
   \vspace{-1cm}
   \caption{
 {\footnotesize{\sl We list the pure spinor constraints and the  RR or NSNS fields to which they are associated. We denote
 by $\lambda_A$ with $A={1,2}$ the pure spinor either in type IIA and type IIB. In the former case $A=1,2$
 are of opposite chirality, while for the latter they have the same chirality. $T$ stands for the
 torsion constraint. The vielbein $V^{\underline{a}}$ is the pull-back on the worldsheet of the target space vielbein.
  }}
 \label{tabellaPS}}
\end{table}
}
 In paper
\cite{marp}, we deduced the pure spinor constraints from the closure of the
FDA algebra in its extended ghost-form by suppressing all the other ghosts  except those
of supersymmetry. It is important to clarify how the constraints in
table \ref{tabellaPS} have to be understood. The latter are too strong
for a 10d target-space vielbein $V^{\underline a}$ and therefore
we have to project them on the 2d string worldsheet  by embedding it into the target-space.
Explicitly  the vielbeins $V^{\underline a}$ must be replaced by their pull-back onto the worldsheet, namely:
\begin{equation}
  V^{\underline{a}} \, \mapsto \, \Pi^{\underline a}_+ \, e^+ \, +\,
  \Pi^{\underline a}_- \, e^-
\label{embeddo}
\end{equation}
where $e^\pm$ denote the worldsheet zweibein.
\par
  In this way, we are able to prove that
the number of independent degrees of freedom is the same in all cases.
\par
The proof of the equivalence is done in two ways. First, we  find the solutions of the pure spinor constraints  using an
$\mathrm{SO(8)}$ and a $\mathrm{G_2}$ decomposition, respectively. Both solutions are quite interesting: the former  provides a
solution in terms of an infinite number of fields (this solution is very similar to the one proposed in \cite{Berkovits:2000nn}); the latter
shares several similarities, since it preserves only the $\mathrm{G_2}$ invariance, but it can be expressed in terms of
a finite number of fields. We provide a complete solution on a single patch and one has
to extend the solution to the entire space as usual.
Then, we prove the equivalence between type IIA and
type IIB pure spinors, by showing that there exists a simple map between the two set of constraints  which
is written in terms of Dirac matrices. Finally  we show the equivalence of the
solutions for the type IIB and the Poincar\'e case by mapping  one solution in the other.
\par
To complete the proof we observe that in the  type IIA case, all constraints can be
cast into a single Lorentz tensor of anti-de Sitter group  $\mathrm{SO(2,10)}$ which can be written
as follows
\begin{equation}\label{12d}
\Lambda \Gamma_{[\Sigma\Xi]} \Lambda =0\,, \hspace{3cm}  \Sigma, \Xi=1, \dots, 12
\end{equation}
and therefore one can skew-diagonalize the constraint by an $\mathrm{SO(2,10)}$ rotation which can be
expressed in terms of a rotation of $\mathrm{Spin(2,10)}$ on the spinors $\Lambda$. Notice that
also the Poincar\'e pure spinor constraints can be rotated in the same way and therefore, if the
matrices have the same rank (i.e. the same number of non-zero skew-eigenvalues) then we
can map the Poincar\'e constraints into the type IIA ones. Indeed, it is easy to show that the rank of the
pure spinor matrix is four in all cases so that there are two independent sets of skew-eigenvalues.
Thus, by a $\mathrm{Spin(2,10)}$ transformation, we can rotate one type of constraints into the others.
\par
The paper is organized as follows: sec. 2 provides an extract of the derivation of the pure spinor constraints from the FDA of supergravity.
In sec. 3 we discuss the solution of Poincar\'e pure spinor constraints by means of a $\mathrm {G_2}$ decomposition. In sec. 4,
we compute the number of independent components
using either a $\mathrm {G_2}$ or an $\mathrm{SO(8)}$ decomposition. In sec. 5, we derive the equations of pure spinors for the
type IIA case in the $\mathrm {G_2}$ decomposition and finally in sec. 6 we prove the equivalence.
\section{Pure Spinor Constraints from FDA}
\label{Psconstra}

Using the same argument used in paper \cite{marp} for the type IIA case, the pure spinor constraints
implied by the constrained ghost extension of the type IIB FDA curvatures are
as follows:
\begin{eqnarray}
{\rm i} \, \overline{\lambda} \, \Gamma^{\underline{a}} \, \lambda & \approx & 0 \label{torsionconst}\\
\Lambda^\eta_+ \, \overline{\lambda} \, \Gamma_{\underline{a}} \, \lambda^\star \, V^{\underline{a}} \, +
\Lambda^\eta_- \, \overline{\lambda^\star}
\, \Gamma_{\underline{a}} \, \lambda \, V^{\underline{a}} \, & \approx &
0\,, \hspace{1cm} \eta = 1,2 \label{A2const}\\
\overline{\lambda} \, \Gamma_{\underline{abc}} \, \lambda \, V^{\underline{a}} \,
\wedge \,V^{\underline{b}} \, \wedge \, V^{\underline{c}} & \approx & 0 \label{C4const}
\end{eqnarray}
where the $2 \times 2$ complex matrix $\Lambda$
denotes the coset representative of $\mathrm{SU(1,1)/U(1)}$, by
$V^{\underline{a}}$ we denote the 10d vielbein and $\lambda$ is a
Weyl commuting spinor.
Eq.(\ref{torsionconst}) follows from the BRST variation of
the diffeomorphism ghosts $\xi^a$, when they are set to zero,
eq.(\ref{A2const}) follows from the BRST variation of the
$2$-form ghosts $a_{[2-i,i]}$ when they are set to zero and by the
same token eq.(\ref{C4const}) follows from the BRST variation of the
$4$--form ghosts.
Only the first of these equations is background independent. On the contrary the second and the third depend also
on the vielbein, namely on the background. The constraint
(\ref{C4const}) has no projection on the string-worldvolume and hence
no relevance. The constraint (\ref{A2const}) instead, written in real
notation looks like follows:
\begin{equation}
  d^{A|BC} \, \overline{\lambda}_{B} \, \Gamma_{\underline{a}} \,
  \lambda_{C} \, V^{\underline{a}} \, = \, 0
\label{verocostretto}
\end{equation}
(where $A,B,C =1,2$)
so that the constraints to be solved are:
\begin{eqnarray}
0  =  \overline{\lambda}_{A} \, \Gamma_{\underline{a}} \,
  \lambda_{A} \, V^{\underline{a}} \,,\hspace{2cm}
0 =  d^{A|BC} \, \overline{\lambda}_{B} \, \Gamma_{\underline{a}} \,
  \lambda_{C} \, V^{\underline{a}}
\label{TrueStory}
\end{eqnarray}
where the $d^{A|BC}$ tensor denotes the Clebsh-Gordon coefficients for the decomposition
of two doublets of $\mathrm{SO(2)}$ into one doublet. So that, $d^{1|BC} \propto (\sigma^3)^{BC}$ and 
$d^{2|BC} \propto (\sigma^1)^{BC}$. 

The structure of the pure spinor constraints in the presence of non trivial backgrounds
were also obtained in \cite{Anguelova:2003sn} where the couplings with RR fields
are essential for D-brane actions.

\section{Poincar\'e Pure Spinors}
\label{critica1}

The Poincar\'e constraints used by Berkovits are rather different from ours.
They are encoded in the following background independent equations
\begin{eqnarray}
\sum_{A=1,2}\overline{\lambda}_A \, \Gamma^{\underline{a}} \, \lambda_A & = & 0\,, \hspace{3cm}
\sum_{A=1,2}(-)^A\, \overline{\lambda}_A \, \Gamma^{\underline{a}} \, \lambda_A  =  0
\label{purespinorconstraints}
\end{eqnarray}
\par
 Hence in this case  we have to solve only the constraint:
\begin{equation}
   \overline{\lambda} \, \Gamma^{\underline{a}} \, \lambda \, = \,0
\label{costrettone}
\end{equation}
where $\lambda$ is a chiral spinor $\Gamma_{11} \, \lambda \, =
\lambda$. We use the gamma matrix basis well adapted to the $1$-brane case
which we describe in the appendix and we solve the $d=10$ chirality condition by posing:
\begin{equation}
  \lambda \, = \, \phi_+ \, \otimes\, \zeta_+ \, + \, \phi_-\,
  \otimes \zeta_-
\label{lambdapongo}
\end{equation}
where $\phi_\pm$ are $2$-component $\mathrm{SO(1,1)}$ chiral spinors and
$\zeta_\pm$ are $16$-components $\mathrm{SO(8)}$ spinors also chiral:
\begin{equation}
  \gamma_3 \, \phi_\pm \, = \, \pm \phi_\pm \quad ; \quad T_9 \,
  \zeta_\pm \, = \, \pm \, \zeta_\pm
\label{chiralspini}
\end{equation}
In the chosen basis we have:
\begin{equation}
  \begin{array}{ccccccc}
    \phi_+ & = & \left( \begin{array}{c}
      \varphi_+ \\
      0 \\
    \end{array}\right)  & ; & \phi_- & = &\left(  \begin{array}{c}
      0 \\
      \varphi_- \\
    \end{array} \right) \\
    \null & \null & \null & \null & \null & \null & \null \\
    \zeta_+ & = & \left( \begin{array}{c}
      0 \\
      \omega_+ \\
    \end{array}\right)  & ; & \zeta_- & = &\left(  \begin{array}{c}
      \omega_- \\
      0 \\
    \end{array} \right) \
  \end{array}
\label{spinorini}
\end{equation}
where $\varphi_\pm$ are just complex numbers while $\omega_\pm$ are
$8$-components complex $\mathrm{SO(7)}$ spinors.
\par
In view of the charge conjugation matrix (\ref{chargconjval}) the
pure spinor constraint (\ref{costrettone}) reduces to:
\begin{equation}
 0 \, = \,  \lambda^T \, C \, \Gamma^{\underline a} \, \lambda \, = \, \left \{\begin{array}{c}
   \phi_+^T \, \epsilon \, \gamma_i \, \phi_+ \, \zeta_+^T \, \zeta_+ \, + \, \phi_-^T \, \epsilon \, \gamma_i \, \phi_- \, \zeta_-^T \, \zeta_- \, = \, 0\quad (i=0,1)\\
  \phi_+^T \, \epsilon \, \phi_- \, \zeta_+^T \, T_I \, \zeta_- \, = \, 0 \quad (I=1,\dots , 8) \
 \end{array}
 \right.
\label{vincoletti}
\end{equation}
which further reduces to:
\begin{eqnarray}
0 & = & \varphi_+^2 \, \zeta_+^T \, \zeta_+ \, - \, \varphi_-^2 \, \zeta_-^T \, \zeta_-  \nonumber\\
0 & = & - \varphi_+^2 \, \zeta_+^T \, \zeta_+ \, - \, \varphi_-^2 \, \zeta_-^T \, \zeta_-  \nonumber\\
0 & = & 2 \, \varphi_+ \, \varphi_- \, \zeta_+^T \, T_I \, \zeta_-
\label{genuina}
\end{eqnarray}
Since $\varphi_\pm$ are $1$-component objects, namely complex numbers,
eq.s (\ref{vincoletti}) yield:
\begin{equation}
  \left. \begin{array}{rcl}
    \zeta_\pm ^T \, \zeta_\pm & = & 0 \\
    \zeta_+ \, T_I \, \zeta_- & = & 0\
  \end{array} \right \} \, \Rightarrow \, \left \{ \begin{array}{rcl}
  \omega_\pm ^T \, \omega_\pm & = & 0 \\
  \omega_+ ^T \, \tau^\alpha \, \omega_- & = & 0 \\
   \omega_+ ^T \, \omega_- & = & 0 \
  \end{array}\right.
\label{tettona}
\end{equation}
The constraints in eq.(\ref{tettona}) can be solved in various ways.
We have four branches of a singular solution depending only on $7$
complex parameters and a regular solution depending on $11$ complex
parameters.

{\bf The singular solution with $7$-parameters}
Let $\omega=\{\omega_1 , \dots ,\omega_8 \}$ be an $8$-component complex spinor
fulfilling the equation
\begin{equation}
  \omega^T \, \omega \, = \, 0
\label{omsq=0}
\end{equation}
then eq.s(\ref{tettona}) can be solved by setting either:
\begin{description}
  \item[1)] $\omega_+ \, = \, \omega \quad ; \quad \omega_- \, = \, 0 \quad$
  or
  \item[2)] $\omega_+ \, = \, 0 \quad ; \quad \omega_- \, = \, \omega \quad$
  or
  \item[3)] $\omega_+ \, = \, \, \omega \quad ; \quad \omega_- \, = \, \omega\quad $
  or
  \item[4)] $\omega_+ \, = \, \, \omega \quad ; \quad \omega_- \, = \, - \, \omega$
\end{description}

{\bf The regular solution with 11 parameters}
Let $\varpi^\alpha$ and $\chi^\alpha$ be two $7$-components complex vectors
satisfying the constraints:
\begin{eqnarray}
\varpi \, \cdot \, \varpi \, \equiv\, \varpi^\alpha \, \varpi^\alpha& = & 0 \nonumber\\
\varpi \, \cdot \, \chi \, \equiv\, \varpi^\alpha \, \chi^\alpha& = & 0
\label{ortogonalini}
\end{eqnarray}
We can solve eq.s(\ref{tettona}) setting either
\begin{description}
  \item[1)]
\begin{equation}
  \begin{array}{rclcrcl}
    \omega_+^\alpha & = & \varpi^\alpha & ; & \omega^8_+ & = & 0 \\
    \omega^\alpha_- & = & a^{\alpha \beta \gamma }\, \varpi^\beta \, \chi^\gamma & ; & \omega^8_- & = & 0 \
  \end{array}
\label{1branco}
\end{equation}
or
  \item[2)]
  \begin{equation}
  \begin{array}{rclcrcl}
  \omega^\alpha_+ & = & a^{\alpha \beta \gamma }\, \varpi^\beta \, \chi^\gamma & ; & \omega^8_+ & = & 0
  \\
    \omega_-^\alpha & = & \varpi^\alpha & ; & \omega^8_- & = & 0 \\
    \end{array}
\label{2branco}
\end{equation}
\end{description}
In this way we have $11$ eleven parameters in each of the two pure
spinors $\lambda_1$ and $\lambda_2$. Indeed $\varpi^\alpha$ counts
for $6$ because its norm is zero and $\pi^\alpha$ counts for $5$
because it is orthogonal to a vector of vanishing norm and because it
is defined up to a gauge transformation $\chi^\alpha \, \mapsto \,
\chi^\alpha \, + \, x \, \varpi^\alpha$ where $x$ is the gauge parameter.
This makes the correct counting $22$.
\par
Together with the ghost fields $\lambda$, one has to consider their conjugate momenta
$w$.
We recall the quadratic part of the action for free pure spinors (here we consider only the left-moving
sector for simplicity)
\begin{equation}\label{acA}
S= \int w_\a \bar\partial \l^\a = \int w^{T} \bar\partial \lambda\,.
\end{equation}
where we have also neglected the coupling with the homolorphic form $\Omega$ (see
for example \cite{Nekrasov:2005wg,Adam:2006bt,Berkovits:2007wz})
since it does not enter in the present discussion.
We have also used the matrix notation $w^T$ to denote the spinor $w_\a$.

We observe that if  $\l^\a$ satisfies the pure spinor constraints, which are first class
constraints (since their Poisson brackets vanish), then there are gauge symmetries generated by
them. If we denote by $q =\oint \Lambda_m \overline\l\Gamma^m  \l$ the charge associated to that gauge symmetry and $\Lambda_m$ a set of gauge parameters, then
we have the gauge transformations
\begin{equation}\label{acB}
\delta w_\a = 2 \, \Lambda_m (C \Gamma^m \l)_\a\,.
\end{equation}
Now, in order to use the decomposition (\ref{lambdapongo}), we insert it
in the above equation and we use the fact that the spinors $\phi_\pm$ have only one non-zero component. Hence, their value can be reabsorbed into $\zeta_\pm$ and they can be set equal to 
unit versors $\phi_+ = (1,0)$ and $\phi_-=(0,1)$,  yielding  
\begin{equation}\label{acC}
S= \int w^T \left( \phi_+ \otimes \bar\partial \zeta_+  + \phi_- \otimes \bar\partial \zeta_- \right)\,.
\end{equation}
Since the spinors $\phi_\pm$ are orthogonal to each other, we can decompose  the
conjugates $w_\a$ as follows
\begin{equation}
w \, = \, \phi_+ \, \otimes \, {\sf w}^- \, + \, \phi_- \, \otimes  {\sf w}^+
\end{equation}
where $ {\sf w}^\pm$ are $8$-dimensional spinors and
the action becomes
\begin{equation}\label{acD}
S=
\int \left( {\sf w}^- \bar\partial \omega_+  + {\sf w}^+ \bar\partial \omega_- \right)  \,.
\end{equation}
Here we have plugged the definitions (\ref{spinorini}). Hence from the pure spinor constraints
written in terms of $\omega_\pm$ (see eq.~(\ref{tettona})),
 it is straightforward to get
 \begin{equation}
 \delta  \wipm = \Lambda^\pm \wipm + \widehat\Lambda \wimp +
\Lambda_\alpha \tau^\alpha \wimp\,,
\end{equation}
where $\Lambda^\pm, \widehat\Lambda$ and $\Lambda_\alpha$ are 10 gauge parameters obtained
by decomposing the vector $\Lambda_m$ in representations $(2,0) + (0,0) + (0,7)$ of $\mathrm{SO(1,1) 
\otimes SO(7)}$ as in (\ref{acB}). The pure spinor constraints are not irreducible and therefore the gauge symmetries are not all independent. It is easy to see that one can use the three gauge parameters $\Lambda^\pm, \widehat\Lambda$ to set some components of $\wipm$
to zero, but the gauge transformation of the $8^{\rm th}$-component  (because of the
ansatz (\ref{2branco})) works as follows
\begin{equation}
\delta \wip_8 = \Lambda_\alpha \varpi^\alpha\,,
\hspace{2cm}
\delta \wim_8 = \Lambda_\alpha a^{\alpha\beta\gamma} \chi_\beta\varpi_\gamma\,.
\end{equation}
This implies that the two components of $\wipm$ can be set to zero by using
the gauge parameters $\Lambda_\alpha$. The independent gauge parameters is the
space complementary to that spanned by the solutions of
\begin{equation}
 \Lambda_\alpha \varpi^\alpha = 0\,,
 \hspace{2cm}
 \Lambda_\alpha a^{\alpha\beta\gamma} \chi_\beta\varpi_\gamma = 0\,.
 \label{piffo}
\end{equation}
It is easy to count the gauge parameters by observing that the two constraints
imply that there are 5 free parameters, defined up to some
the gauge symmetries $\Lambda_\alpha \rightarrow \Lambda_\alpha + x \varpi_\alpha +
y \chi_\alpha + z a_{\alpha\beta\gamma} \chi^\beta \varpi^\gamma$ with $x,y,z$ the corresponding gauge parameters. This yields the wanted gauge parameter counting:  seven parameters $\Lambda_\alpha$ minus two constraints (\ref{piffo}) minus
the three gauge symmetries makes two. Therefore, the number of non-vanishing $w$ can be fixed to eleven. 

\section{PS for IIB backgrounds}
\label{critica2}

\subsection{Solution with $\mathrm{G}_2$ decomposition}
Let us now compute the solution of the pure spinor constraints  in the case of
type IIB backgrounds and and let us show that there is a $22$-parameter solution also for them,
although differently constructed. Also in this case we use a well
adapted basis of gamma matrices and we search for a $\mathrm{G_2}$-covariant
parametrization of the solution.
\par
Let us then consider eq.s (\ref{TrueStory}) and let us treat them as before
by setting the following tensor product parametrization:
\begin{equation}
  \lambda_A \, = \, \phi_+ \, \otimes \, \zeta_A^+ \, + \, \phi_- \,
  \otimes \, \zeta^-_A
\label{tensoproducto}
\end{equation}
where:
\begin{equation}
  \begin{array}{ccccccc}
    \phi_+ & = & \left( \begin{array}{c}
      1 \\
      0 \\
    \end{array}\right)
     & ; & \phi_- & = & \left(\begin{array}{c}
       0 \\
       1 \\
\end{array} \right)  \\
    \zeta_A^+ & =  & \left(\begin{array}{c}
      0 \\
      \omega^+_A \
    \end{array}  \right) & ; & \zeta_A^- & =  & \left( \begin{array}{c}
      \omega_A^- \\
      0 \
    \end{array}\right)  \
  \end{array}
\label{blocchini}
\end{equation}
In writing eq.s(\ref{blocchini}) we have observed that the unique
component of $\phi_\pm$ can always be reabsorbed in the normalization
of $\omega_A^\pm$ and hence set to one, as already noted.
\par
Using a well adapted basis where, defining, $V^{\underline{a}} \, = \, \Pi^{\underline{a}}_i \,
e^i$ we have:
\begin{equation}
  \Pi^{j}_i \, = \, \delta^{j}_i \quad \mbox{and zero otherwise}
\label{piadatto}
\end{equation}
the constraints to be solved reduce to the following ones:
\begin{eqnarray}
\phi_+^T \, \epsilon \, \gamma_i \, \phi_+ \, \zeta_A^+ \, \cdot \, \zeta_A^+ \, + \,
\phi_-^T \, \epsilon \, \gamma_i \, \phi_- \, \zeta_A^- \, \cdot \, \zeta_A^-  =  0
\label{abbaziadinovalesa}
\end{eqnarray}
$$
d^{A|BC} \, \left( \phi_+^T \, \epsilon \, \gamma_i \, \phi_+ \, \zeta_B^+ \, \cdot \, \zeta_C^+ \, + \,
\phi_-^T \, \epsilon \, \gamma_i \, \phi_- \, \zeta_B^- \, \cdot \, \zeta_C^- \right)   = 0
$$
$$
\phi^T_+ \, \epsilon \, \phi_- \, \zeta_A^+ \, \cdot \, T^I \,
\zeta^- =  0\,.
$$
Elaborating eq.s (\ref{abbaziadinovalesa}) a little further we reduce
them to the following ones in terms of $8$-component
$\mathrm{SO(7)}$-spinors:
\begin{eqnarray}
\omega_1^\pm \, \cdot \, \omega^\pm_1 & = & 0 \label{our1}
\label{our5}
\end{eqnarray}
$$\omega_2^\pm \, \cdot \, \omega^\pm_2  =  0 \label{our2}$$
$$\omega_1^\pm \, \cdot \, \omega^\pm_2  =  0 \label{our3}$$
$$\omega_1^+ \, \cdot \, \omega^-_1 \, + \, \omega_2^+ \, \cdot \, \omega^-_2  =  0 \label{our4}$$
$$\omega_1^+ \, \cdot \, \tau^\alpha \, \omega^-_1 \, + \, \omega_2^+ \, \cdot \, \tau^\alpha \, \omega^-_2 =  0$$
It is now easy to present the $22$-parameter solutions of the above
constraints. Let
\begin{equation}
  \varpi^\alpha \quad ; \quad \pi^\alpha \quad ; \quad \xi^\alpha
  \quad ; \quad \chi^\alpha
\label{quattrocose}
\end{equation}
be a set of four $7$-component vectors (fundamental representations
of $\mathrm{G_2}$) subject to the following constraints:
\begin{eqnarray}
  \varpi \, \cdot \, \varpi & = & 0 \label{purga1}\\
\pi \, \cdot \, \pi & = & 0 \label{purga2}\\
a^{\alpha \beta \gamma } \, \chi_\alpha \, \pi_\beta \, \varpi_\gamma
& = & 0 \label{purga3}\\
a^{\alpha \beta \gamma } \, \xi_\alpha \, \pi_\beta \, \varpi_\gamma
& = & 0 \label{purga4}
\end{eqnarray}
the solutions of the constraints (\ref{our1}-\ref{our4}) is given by
the following positions:
\begin{eqnarray}
\omega^+_1 & = & \left(\varpi^\alpha \, , \, 0 \right)  \nonumber\\
\omega^-_2 & = & \left(\pi^\alpha \, , \, 0 \right)  \nonumber\\
\omega^-_1 & = & \left(a^{\alpha \beta \gamma }\, \chi_\beta \, \varpi_\gamma \, , \, \chi\, \cdot \, \varpi
\right)\nonumber\\
\omega^+_2 & = & \left(a^{\alpha \beta \gamma }\, \xi_\beta \, \pi_\gamma \, , \, \xi\, \cdot \, \pi
\right)
\label{soluzia}
\end{eqnarray}
It is easy to count the number of parameters and verify that it
amounts to $22$ independent ones. Indeed $\varpi$ and $\pi$ count $6$
each being of vanishing norm, while $\chi$ and $\xi$ count $5$ each
because they are subject to the constraints
(\ref{purga3}-\ref{purga4}) and defined up to a gauge
transformation. 
Hence we have a $22$-parameter solution also for  the FDA pure
spinor constraints for type IIB. 

The intersection of the this solution with the Poincar\'e solution 
is obtained by imposing the extra condition:
\begin{equation}
  \varpi \, \cdot \, \pi \, = \, 0
\label{intersetta}
\end{equation}
that reduces the space to a $21$-parameter one and does not yield
the correct counting.

Again, we want to check that the conjugate momenta have the correct degrees of freedom also 
in the FDA case. For that we introduce the two sets of $8$-dimensional spinors $\wipm_A$
with $A=1,2$ and we derive their gauge transformations from the reduced
equations (\ref{our5}) to get
\begin{eqnarray}
&&\delta \wipm_1 = 2 \Lambda^\pm_1 \omega^\pm_1 +
\widehat\Lambda^\pm \omega^\pm_2 + \widetilde \Lambda \omega^\mp_1 +
\Lambda_\alpha \tau^\alpha \omega^\mp_1\,, \\
&&\delta \wipm_2 = 2 \Lambda^\pm_2 \omega^\pm_2 +
\widehat\Lambda^\pm \omega^\pm_1 + \widetilde \Lambda \omega^\mp_2 +
\Lambda_\alpha \tau^\alpha \omega^\mp_2\,, \end{eqnarray}
where $\Lambda_A^\pm, \widehat\Lambda^\pm, \widetilde\Lambda$ and $\Lambda_\alpha$ are the gauge parameters. As before, we can use $\Lambda_A^-$ to fix the two $8^{\rm th}$-components of $\wipm_{8 A}$ to zero and we can use the remainng parameters to fix other two components of $\wipm_A$ by a combination of $\Lambda_\alpha$ and
$\widetilde\Lambda$. We are left with $\widehat\Lambda^\pm$ and
$\Lambda^+_A$ two set other fours to zero. Finally, we can use other twos of
$\Lambda_\alpha$ to set the components of $\wipm_A$ to zero. However,
of the original 7 components of $\Lambda_\alpha$, 4 components are irrelevant  and among the 
other three components there are no residual gauge symmetries. Indeed, by setting
to zero the gauge transformation of the $8^{\rm th}$ components of $\wipm_{8 A}$, there is no residual gauge symmetry and we have exactly 3 gauge parameter $\Lambda_\alpha$. 
So, the total counting is again 22. Summarixing also in the FDA case out of the fourteen gauge parameters corresponding to the fourteen pure spinor constraints, only ten are irreducibles and can be used to gauge away ten components of the 32 $\wipm_A$. 

\subsection{Solution with $\mathrm{SO(8)}$ symmetry}

It is convenient to solve the FDA constraints also in a $\mathrm{SO(8)}$ basis. For that we
use the solution for a single pure spinor in an $\mathrm{SO(8)}$ basis of the constraints
\begin{equation}
\label{sottoA}
\omega^\pm_2 \cdot \omega^\pm_2 =0\,, ~~~~~
\omega^+_2 \sigma^I \omega^-_2 =0\,, ~~~~~
\end{equation}
where $\sigma^I$ are the Pauli matrices in eight dimensions and $\omega^\pm_2$ are $8_c$ and $8_s$ representation of $\mathrm{spin}(8)$, respectively. The index $I$ instead runs over $1, \dots, 8$ in the $8_v$ representation. 


To solve (\ref{sottoA}) one can make the ansatz \cite{Berkovits:2000nn} $\omega^-_2 = \eta_I \sigma^I \omega^+_2$ with
$\eta_I$ an eight dimensional vector in the $8_V$ representation. This ansatz solves the
equation $\omega^-_2\cdot \omega^-_2 =0$ and the third equation in (\ref{sottoA}) if
$\omega^+_2 \cdot \omega^+_2 =0$. This implies that there are 7 independent components for $\omega^+_2$ and
$\omega^-_2$ is expressed in terms of $\omega^+_2$ and in terms of $\eta_I$. However, the latter are defined up to an infinite number of gauge degrees of freedom (since we can shift $\eta_I$ with
$\eta_I + v \sigma_I \omega^{+}$ where $v \in 8_s$ and again the latter is defined up to gauge degrees of freedom. This procedure iterates up to infinity) which effectively makes the counting of independent components of $\omega^-_2$ equal to 4. Explicitly, this can be done by breaking $\mathrm{SO(8)}$ 
to $\mathrm{SU(4)}$ and suppressing one of the two four into which the $8_v$ breaks up. So, this sums up to 11 components for $\omega^\pm_2$.

Next, we make the ansatz
\begin{equation}
\label{sottoB}
\omega^\pm_1 = \alpha^\pm \omega^\pm_2 + \alpha^\pm_I \sigma^I \omega^\mp_2\,.
\end{equation}
where $\omega_2^\pm$ solves equation (\ref{sottoA}))
Here $\alpha^\pm$ and $\alpha_I^\pm$ are $2+2\times 8$ independent degrees of freedom
which parametrize the solution for $\omega^\pm_1$ in terms of $\omega^\pm_2$. Notice that
the amount of parameters exceeds the wanted independent components of $\omega_1^\pm$. This
means that we have to reduce the number of them by imposing some relations. Indeed, by inserting
the ansatz (\ref{sottoB}) into (\ref{our5}), and using (\ref{sottoA}),
we get that $\alpha^\pm$ are free independent parameters, but $\alpha^\pm_I$ are constrained by
\begin{equation}
\label{sottoC}
\alpha^+_I \alpha^-_J - \alpha^+_J \alpha^-_I =0\,.
\end{equation}
To derive (\ref{sottoC}) one has to use the commutation relations and the symmetry properties of the products of the Pauli matrices
and using the fact that $\omega^\pm_2$ are bosonic quantities. The most general solution of
(\ref{sottoC}) has indeed $8+1$ parameters. Then, the total independent parameters which
describe the solution for $\omega_1^\pm$ are effectively 11, which is the correct counting. So, the
solution is asymmetric, but it takes into account the $\mathrm{SO(8)}$ symmetry.

In order to compare this solution of the type FDA constraints with a symmetric solution of the
Poincar\'e ones we can observe that the asymmetric solutions for $\omega^\pm_1$ is written
in terms of $\omega^\pm_2$. On a patch where $\a^+ \neq 0$ we can solve $\omega^+_2$
in terms of $\omega^+_1$ yielding
\begin{equation}\label{relA}
\omega^+_2 = \frac{1}{\a^+}(\omega^+_1 + \a_I^+ \sigma^I \omega^-_2)\,.
 \end{equation}
Inserting this result into $\omega^-_1$, one gets
\begin{equation}
\omega^-_1 = \a^- \omega^-_2 +
\frac{1}{\a^+} \a_I^+ \sigma^I \omega^+_1
+ \frac{1}{\a^+} \a_J^-  \a_I^+ \sigma^J \sigma^I \omega^-_2)
\end{equation}
$$
= \a^- \omega^-_2 +
\frac{1}{\a^+} \a_I^+ \sigma^I \omega^+_1
+ \frac{1}{2 \a^+} \Big( \a_J^-  \a_I^+ \sigma^J \sigma^I +
\a_I^-  \a_J^+ \sigma^I \sigma^J \Big)
\omega^-_2
$$
$$
=\Big(\a^- - \frac{1}{\a^+} \a_I^- \a^+_J \delta^{IJ}\Big) \omega^-_2 +
\frac{1}{\a^+} \a_I^+ \sigma^I \omega^+_1
$$
So, finally if we choose to have $\a^+ \a^- = \a_I^- \a^+_J \delta^{IJ}$ (notice that
this equation is again a cone) we get $\omega^-_1 = \hat \a_I^- \sigma^I \omega^+_1$ which has the
form of the solution for $\omega^-_2$ and it is symmetric. Notice that we have chosen $\alpha^-$ to 
put the solution in the wanted form and this reduces the amount of independent dof to 21. For 
a generic solution of Poincar\'e pure spinor, one has to uplift the constraints on $\alpha$'s in order 
to satisfy the symmetric constraints. Also in $\mathrm{SO(8)}$ we find that the intersection space of solutions has 21 parameters. 

So, we have found a map between the solution for type FDA type IIB pure 
spinors to the symmetric solution of the Poincar\'e constraints.

\section{FDA PS for IIA backgrounds}
\label{critica3}

Let us now compute the solution of the pure spinor constraints in the case of
type IIA backgrounds and let us show that there is a $22$-parameter solution also for them,
although differently constructed. Also in this case we use a well
adapted basis of gamma matrices and we search for a $\mathrm{G_2}$ invariant
parametrization of the solution. We show that reducing the pure spinor constraints to the
$\mathrm{G_2}$ basis, one gets the same equations as in (\ref{our1}).
\par
As displayed in table \ref{tabellaPS}, we have the following PS constraints:
\begin{eqnarray}
0&=&\sum_A \overline\lambda_A \Gamma_{\underline{a}} \lambda_A \,, \hspace{3cm}
0=\sum_A (-)^I \overline\lambda_A \Gamma_{\underline{a}} \lambda_A  V^{\underline{a}}\,, ~~~
\label{IIAstory}\\
0&=&\overline\lambda_1 \Gamma_{[{\underline{a}} {\underline{b}}]} \lambda_2
       V^{\underline{a}} V^{\underline{b}} \,, \hspace{3cm}
0=\overline\lambda_1 \lambda_2 =0\,.
\label{IIAstoryA}
\end{eqnarray}
The first two constrains come from the torsion and from the $\mathbf{B}_{[2]}$ form, whilst the other
two are coming from the variation of the RR fields $\mathbf{C}_{[1]}$ and $\mathbf{C}_{[3]}$.

As above, we write the PS $\lambda_I$ by decomposing them using the same basis and we
get the two structures
\begin{eqnarray}
&&  \lambda_1 \, = \, \phi_+ \, \otimes \, \zeta_1^+ \, + \, \phi_- \,
  \otimes \, \zeta^-_1 \\
&&  \lambda_2 \, = \, \phi_+ \, \otimes \, \zeta_2^- \, + \, \phi_- \,
  \otimes \, \zeta^+_2 \,,
  \label{IIAstoryB}
\end{eqnarray}
and inserting them into (\ref{IIAstory})-(\ref{IIAstoryA}) we get the following equations
\begin{eqnarray}
0&=& \phi^T_+ \epsilon \gamma_i \phi_+ \left( \zeta_1^+ \zeta_1^+ \pm \zeta_2^- \zeta_2^- \right) +
\phi^T_- \epsilon \gamma_i \phi_- \left( \zeta_1^- \zeta_1^- \pm \zeta_2^+ \zeta_2^+ \right) \,, \nonumber
\\
0&=& \phi^T_+ \epsilon \phi_+ \left( \zeta_1^+ T^I \zeta_1^- + \zeta_2^-  T^I
\zeta_2^+ \right) \,, \nonumber \\
0&=& \phi^T_+ \epsilon \phi_+ \left( \zeta_1^+ \zeta_2^+ - \zeta_1^-
\zeta_2^- \right) \,, \nonumber \\
0&=& \phi^T_+ \epsilon \gamma_{+-} \phi_+ \left( \zeta_1^+ \zeta_2^+ + \zeta_1^-
\zeta_2^- \right) \,,
\label{IIAnew1}
\end{eqnarray}
which completely  reduce to eq.s (\ref{our1}). This shows that reducing the equations to
the present $\mathrm{G}_2$ basis, one can verify the T-duality of the FDA PS constraints. It follows that
the solutions are also isomorphic even though the set of constraints are different. We conclude that
all three sets of constraints are equivalent even though the solutions differ (but with the same number of parameters). 

\subsection{From IIB to IIA FDA PS}

In order to map the pure spinor constraints of tyoe IIA  to those of type IIB, we consider the following map
\begin{equation}\label{mapA}
\l_2 \rightarrow \Big(\alpha \Gamma^+ + \beta \Gamma^- \Big)\l_2\,.
\end{equation}
Inserting this map into the two constraints $\l_1 \l_2 =0$ and $\l_1 \Gamma^{+-} \l_2 =0$
using an adapted basis, we get 
\begin{eqnarray}\label{mapB}
&&\alpha \l_1 \Gamma^+ \l_2 + \beta \l_1
\Gamma^- \l_2 =0\,, \\
&& \alpha \l_1 \Gamma^+ \l_2 =0\,
\end{eqnarray}
and if $\alpha, \beta \neq 0$, they imply the pure spinor constraints of the type IIB FDA (in an adapted basis).
Next equations, we study the constraint coming form the torsion and from the NS-NS 2-form. 
They read, in an adapted basis,
as follows $\l_1 \Gamma^\pm \l_1 =0$ and $\l_2 \Gamma^\pm \l_2=0$. Recalling that $(\Gamma^\pm)^T = \Gamma^\mp$ we get that using the map (\ref{mapA}), these constraints are mapped into each other. The remaining constraints are easily shown to be equivalent. This completes the proof that the FDA type IIA and type IIB constraints are equivalent.

\section{Overall Equivalence}

In the previous sections we have shown that Poincar\'e and FDA constraints have solutions with the 
same number of parameters which can be mapped into each other. This suggests that the very 
constraint equations are equivalent in the sense that they can be mapped one into the others. This 
is precisely what we are proving in this section. 

To this end, we observe that the FDA type IIA constraints can be organized as follows
\begin{equation}\label{equiA}
\Lambda \Gamma_{\Sigma,\Xi} \Lambda F^{\Sigma,\Xi}_{\Omega \Delta} =0
\end{equation}
where $\Lambda$ is the spinor obtained by combining the two chiral spinors $\l_A$ and $\Gamma_{\Sigma,\Xi}$ are
Dirac matrices of $\mathrm{SO(2,10)}$. The indices $\Sigma,\Xi,\Omega,\Delta$ run over $1,\dots,12$.
Explicitly we have
\begin{equation}\label{equiB}
\left(
\begin{array}{ccccc}
0 & \Lambda \Gamma_{1,2} \Lambda  & \dots &  \Lambda \Gamma_{1, 11}  \Lambda & \Lambda \Gamma_{1}  \Lambda   \\
-\Lambda \Gamma_{1,2} \Lambda & 0  & \dots &  \Lambda \Gamma_{2, 11}  \Lambda  & \Lambda \Gamma_{2}  \Lambda   \\
\vdots & \vdots & \dots & \vdots  & \vdots \\
-\Lambda \Gamma_{1 ,11} \Lambda &  -\Lambda \Gamma_{2, 11}  \Lambda   & \dots  &  0  & \Lambda \Gamma_{11} \Lambda   \\
-\Lambda \Gamma_{1} \Lambda &  -\Lambda \Gamma_{2}  \Lambda   & \dots  &   -\Lambda \Gamma_{11} \Lambda  & 0   \\
\end{array}
\right)
\end{equation}
where for simplicity we have used the indices from $1$ to $10$ for the vectors in 10d and $\Gamma_{\Sigma,12} = \Gamma_\Sigma$ and
$\Gamma_{12,12}=0$.
The matrix (\ref{equiB}) is antisymmetric and therefore it can be skew-diagonalized by an $\mathrm{SO(2,10)}$ rotation. By the invariant theory,
an antisymmetric matrix has only the skew-eigenvalues as invariants and therefore if two matrices have the same number of
eigenvalues (same rank) they are equivalent. Once we have established the rank and found the rotation ${\cal R}$ of $\mathrm{SO(2,10)}$, we
can find the corresponding rotation on the spinors $\Lambda$ as a rotation of $\mathrm{Spin}(32)$. 
After recalling this fact, it is straightforward to verify that both $F^{\Sigma,\Xi}_{\Omega \Delta}$ representing the FDA and the Poincar\'e case have only two non-vanishing skew-eigenvalues. Hence 
the same rank. Indeed, the set of constraints (\ref{equiA})
are viewed as 12d covariant constraints and therefore the rotations of them can be achieved by a $\mathrm{Spin}(32)$ rotation on the spinors $\Lambda$. This completes the proof.

\section*{Remarks and Conclusions}
\begin{itemize}
\item This analysis implies that the canonical form of supergravity as formulated  in the FDA approach 
and that corresponding to the unconventional superspace constraints derived from the 
pure spinor formulation given in \cite{Berkovits:2001ue} are related by "superconformal" transormation 
of $\mathrm{SO(2,10)}$. This also confirms the results of our work \cite{marp}. 

\item As already pointed out,  the Poincar\'e constraints are background independent and the solution \
it does not depend upon the point on the base manifold. In our case, the FDA constraints are 
soldered on the base manifold and therefore, one has to choose an adapted basis to solve. However, 
since the spirit of the entire constructions is to avoind solving them, actually there is no practical difference. On the other hand, the advantage of the FDA approach to pure spinor is the conventional 
framework for the supergravity and therefore yields an explicit 
recipe for the construction of the pure spinor sigma model on any supergravity background. 

\item In the case of heterotic sigma model, the FDA pure spinor constraints coincide with 
the Poincar\'e ones as been noticed by \cite{Oda:2001zm}. 

\end{itemize}

\section*{Acknowledgments}
We thank R. D'Auria, Y. Oz, D. Sorokin, G. Policastro, M. Tonin, M. Trigiante and P. Vanhove for very useful discussions and comments. 

\subsection*{Appendix A: $D=1+9$ basis of gamma matrices well adapted to
$10=1+1\oplus8$}
\label{1+1+8}
In the discussion of the BRST invariant string action and in order to
prove relevant Fierz identities we need to use a different  basis of gamma matrices,
well adapted to the subalgebra:
\begin{equation}
  \mathrm{SO(1,1)} \oplus \mathrm{SO(8)} \, \subset \, \mathrm{SO(1,9)}
\label{11+8}
\end{equation}
We obtain a $32 \times 32$ realization of the $\mathrm{SO(1,9)}$ Clifford
algebra by writing:
\begin{equation}
  \Gamma_{\underline{a}} \, = \,\left \{\begin{array}{rclcl}
    \Gamma_{{i}} & = & \gamma_i \, \otimes \, T_9 &; & i=0,1,\\
    \Gamma_{{1+\Lambda}} & = & \mathbf{1} \, \otimes \, T_I &; & I=1,2,\dots,8\
  \end{array} \right.
\label{gammaconstruzia}
\end{equation}
where $\gamma_i$ are $2 \times 2 $ gamma matrices for the $\mathrm{SO(1,1)}$
Clifford algebra, namely:
\begin{equation}
  \left\{ \gamma_i \, , \, \gamma_j\right\} \, = \, 2 \, \eta_{ij} \,
  = \, \mbox{diag}\left\{ +,-\right\}
\label{gammapiccole}
\end{equation}
while $T_I$ are $16 \times 16$ gamma matrices for the $\mathrm{SO(8)}$
Clifford algebra with negative metric:
\begin{equation}
\left\{ T_I \, , \, T_J\right\} \, = - \, 2 \, \delta_{IJ}
\label{Tamatrici}
\end{equation}
As an explicit representation of the $d=2$ gamma matrices we can take the following ones in terms of Pauli
matrices:
\begin{equation}
  \gamma_0 \, = \, \sigma_1 \, = \, \left( \begin{array}{cc}
    0 & 1 \\
    1 & 0 \
  \end{array}\right) \,, \quad \gamma_1 \, = \, {\rm i} \,
  \sigma_2 \, = \, \left(\begin{array}{cc}
    0 & 1 \\
    -1 & 0 \
  \end{array} \right) \,, \quad \gamma_3 \, = \,
  \sigma_3 \, = \, \left(\begin{array}{cc}
    1 & 0 \\
   0 & -1 \
  \end{array} \right)
\label{gammine}
\end{equation}
On the other hand the $\mathrm{SO(8)}$ Clifford algebra with negative metric
admits a representation in terms of completely real and antisymmetric matrices. We
adopt the following one:
\begin{equation}
  T_{I} \, = \,\left \{\begin{array}{rclcl}
    T_{\alpha} & = & \sigma_1 \, \otimes \, \tau_\alpha &; & \alpha=1,2,\dots,7 \\
    T_8 & = & {\rm i} \, \sigma_2 \, \otimes \, \mathbf{1}_{8 \times 8} &; &
    \null
  \end{array} \right.
\label{Tmatricione}
\end{equation}
where $\tau_\alpha$ denotes the $8 \times 8$ completely antisymmetric realization
of the $\mathrm{SO(7)}$ Clifford algebra with negative metric:
\begin{equation}
  \left\{ \tau_\alpha \, , \, \tau_\beta \right\} \, = - \, 2 \, \delta_{\alpha\beta}
  \, \quad ; \quad \tau_\alpha \, = \, - \left( \tau_\alpha \right)^T
\label{taupiccole}
\end{equation}
given by:
\begin{equation}
  \left( \tau_{\alpha} \right) _{\beta\gamma} = a_{\alpha\beta\gamma}
   \quad ; \quad \left( \tau_{\alpha} \right) _{\beta 8} = - \left( \tau_{\alpha} \right) _{8
   \beta} \, = \, \delta_{\alpha\beta}
\label{ortofresco}
\end{equation}
where the completely antisymmetric tensor $a_{\alpha\beta\gamma}$
encodes the  structure constants of the octionon algebra or, equivalently corresponds to the components of the unique $\mathrm{G_2}$
invariant $3$--form. Explicitly the tensor $a_{\alpha\beta\gamma}$ is
defined by its seven non vanishing components:
\begin{equation}
  \begin{array}{rclcrcl}
    a_{123} & = & -1 & ; & a_{136} & = & -1 \\
    a_{145} & = & -1 & ; & a_{235} & = & -1 \\
    a_{246} & = & 1 & ; & a_{347} & = & -1 \\
    a_{567} & = & -1 & ; & \null & \null & \mbox{all other vanish} \
  \end{array}
\label{atensor}
\end{equation}
The tensor $a_{\alpha\beta\gamma}$ satisfies the following identity:
\begin{equation}
  a_{\alpha\beta\gamma} \,a_{\delta \eta \gamma} = \delta_{\alpha \delta
  } \, \delta_{\beta \eta } \, - \, \delta_{\alpha \eta
  } \, \delta_{\beta \delta } \, - \, \widetilde{a}_{\alpha \beta \delta \eta }
\label{identitus}
\end{equation}
where the complete antisymmetric $4$-index tensor $\widetilde{a}_{\alpha \beta \delta \eta }$ is the dual of
$a_{\alpha\beta\gamma}$. Its non vanishing components are the
following ones:
\begin{equation}
  \begin{array}{rclcrcl}
    \widetilde{a}_{1234} & = & -1 & ; & \widetilde{a}_{1357} & = & 1 \\
    \widetilde{a}_{1256} & = & -1 & ; & \widetilde{a}_{1467} & = & -1 \\
    \widetilde{a}_{2367} & = & -1 & ; & \widetilde{a}_{2457} & = & -1 \\
    \widetilde{a}_{3456} & = & -1 & ; & \null & \null & \mbox{all other vanish} \
  \end{array}
\label{tildeatensor}
\end{equation}
Finally the $16 \times 16$ matrix $T_9$ which anticommutes with all the
$T_A$ has, in this basis, the following structure:
\begin{equation}
  T_9 = - \sigma_3 \, \otimes \, \mathbf{1}_{8\times 8}
\label{T9matra}
\end{equation}
The charge conjugation matrix, with respect to which we have:
\begin{equation}
  C \, \Gamma_{\underline{a}} \, C^{-1} \, = \, - \Gamma_{\underline{a}}^T
\label{chargeconjdef}
\end{equation}
is given by:
\begin{equation}
  C \, = \, \varepsilon \,  \otimes \, \mathbf{1}_{16 \times
  16}\quad ; \quad \left( \varepsilon \, \equiv \, {\rm i} \, \sigma_2
  \,\right)
\label{chargconjval}
\end{equation}

In the paper, we use also the notation $\sigma^I$ for the $8\times8$ Dirac matrices
for the blockdiagonal pieces of $T_I$. Notice that $\sigma^1 = \pm {\bf 1}_{8 \times 8}$  depending
on the chirality.

\end{document}